\documentclass[aps,prl,amsmath,amssymb,twocolumn,superscriptaddress,showpacs]{revtex4}
\usepackage{amsmath,graphicx,float,epsfig,float,tabularx,multirow}
\usepackage{epstopdf,color,bbm,times}

\newcommand{\ket}[1]{| #1 \rangle}
\newcommand{\bra}[1]{\langle #1 |}
\newcommand{\Tr}[1]{\hbox{Tr}\left[ #1 \right]}
\newcommand{\ketbra}[2]{| #1 \rangle \langle #2 |}

\def\hrho{\hat{\varrho}}

\def\e{{\rm e}}
\def\imm{{\rm i}}
\def\Ppnr{{P_{\rm PNR}}}

\def\Npnr{{{\cal N}_{\rm PNR}}}

\begin{document}
\title{Real-time phase-reference monitoring of a quasi-optimal coherent-state receiver}
\author{Matteo Bina}
\affiliation{Dipartimento di Fisica, Universit\`a degli Studi di
Milano, 20133 Milano, Italy}
\affiliation{CNISM UdR Milano Statale, via Celoria 16, 20133 Milano, Italy}
\author{Alessia Allevi}
\affiliation{Dipartimento di Scienza e Alta Tecnologia, Universit\`a degli Studi dell'Insubria, Via Valleggio 11, 22100 Como, Italy}
\affiliation{CNISM UdR Como, via Valleggio 11, 22100 Como, Italy}
\author{Maria Bondani}
\affiliation{Istituto di Fotonica e Nanotecnologie, CNR, Via Valleggio 11, 22100 Como, Italy}
\affiliation{CNISM UdR Como, via Valleggio 11, 22100 Como, Italy}
\author{Stefano Olivares}
\email{stefano.olivares@fisica.unimi.it}
\affiliation{Dipartimento di Fisica, Universit\`a degli Studi di
Milano, 20133 Milano, Italy}
	\affiliation{CNISM UdR Milano Statale, via Celoria 16, 20133 Milano, Italy}
\date{\today}
\begin{abstract}
The Kennedy-like receiver is a quasi-optimal receiver employed in binary phase-shift-keyed communication schemes with coherent states. It is based on the interference of the two signals encoding the message with a reference local oscillator and on/off  photodetection. We show both theoretically and experimentally that it is possible to extract useful information about the phase reference by Bayesian processing of the very same data sample used to discriminate the signals shot by shot. We demonstrate that the minimum uncertainty in phase estimation, given by the inverse of the Fisher information associated with the statistics of the collected data, can be achieved. We also numerically and experimentally investigate the performances of our phase-estimation method in the presence of phase noise, when either on/off or photon-number resolving detectors are employed.
\end{abstract}
\maketitle

In  a phase-estimation protocol, the probe signal is prepared in an optimized pure state that undergoes an unknown phase shift. The phase-shifted signal is then sent to a receiver that retrieves the information about the phase by implementing a suitable detection scheme \cite{hel:76}. The main goal is to reach the minimum uncertainty allowed by the setup, that is related to the Fisher information (FI) or, in the best case, the minimum uncertainty allowed by the very laws of quantum mechanics, which is given by the inverse of the quantum Fisher information \cite{paris:QT}.
\par
On the other hand, in a binary phase-shift-keyed communication (BPSK) channel based on coherent states, a $\pi$ phase shift is imposed or not to an input coherent state $\ket{\beta}$, thus encoding the logical bit ``1'' or ``0''  into the state $\ket{\beta}$ or $\ket{-\beta}$, respectively. In such a case, the problem is turned from the estimation of the phase to the discrimination between the two nonorthogonal phase-shifted coherent states $\ket{\beta}$ and $\ket{-\beta}$ and the goal becomes to reach the minimum error probability in the discrimination process, that is the Helstrom bound \cite{hel:76}.
During the last decade, many solutions, based either on homodyne detection, photon-number resolving (PNR) detectors or hybrid receivers have been theoretically \cite{olivares04,cariolaro10,assalini11} and experimentally proposed \cite{cook07,wittmann10a,wittmann10b,muller12a,izumi:12,becerra:13}.
\par
The typical scheme used to discriminate among two or more optical coherent states involves an apparatus in which the signal interferes at a beam splitter (BS) with a reference coherent state, the local oscillator (LO), whose phase should be known and well defined. Therefore, the crucial point is the control of the phase of the LO \cite{muller12b} and one of the main limitations in the realization of this kind of receivers comes from phase-noise sources, which can affect the generation and propagation of the signals \cite{genoni11, olivares:13}. It is worth noting that the discrimination must be carried out shot by shot, without any \emph{a-priori} knowledge about the transmitted signal state. Such a situation requires the assumption that the relative phase between signal and LO always remains fixed. Moreover, in order to monitor the relative phase in this transmission scheme, the communication protocol must be interrupted and a (known) probe state must be sent to estimate the phase minimizing its uncertainty.
\par
Motivated by the interest on coherent state setups as a resource for deep-space communication \cite{lau:06},  in this Letter we investigate whether and at which extent it is possible to retrieve some information about the relative phase between signal and LO without interrupting the communication, but performing a suitable real-time processing of the same data used for the state discrimination. To this aim, we focus on a simple Kennedy-like receiver \cite{ken73}. Since in its original scheme this kind of receiver employs on/off photodetectors, in the following we will refer to it as on/off receiver. In the present work, we also consider an enhanced version of the receiver, equipped with photon-number resolving detectors (PNR). We show that a suitable analysis of the detector output allows monitoring the phase of the LO down to the minimum uncertainty allowed by the detection scheme.
\par
\textit{Detection scheme} -- Without loss of generality, we assume that the two coherent signals $\ket{\pm \beta}$ considered in the BPSK communication scheme are sent with the same prior probability, $z_0=z_1=1/2$. The overall state reaching the receiver can be then described by the following density operator:
\begin{equation}\label{input}
\hrho(\beta)=\frac12\left(\ketbra{\beta}{\beta} + \ketbra{-\beta}{-\beta}\right).
\end{equation}
The state in Eq.~(\ref{input}) represents a phase-sensitive statistical mixture of two coherent states. To achieve shot-by-shot quasi-optimal discrimination \cite{olivares04} with a Kennedy-like receiver, the state $\hrho(\beta)$ is mixed at a BS of transmittance $\tau$ with a LO excited in the coherent state $\ket{\alpha}$.
\par
By setting $\alpha = \beta\sqrt{\tau/(1-\tau)}$, the overall output state can be written as $\hrho_{\text{out}}(\beta)=\hat{D}(\beta \sqrt{\tau})\hrho(\beta \sqrt{\tau})\hat{D}^\dag (\beta \sqrt{\tau})$, where $\hat{D}(z) = \exp(z \hat{a}^{\dag} - z^* \hat{a})$ is the displacement operator and $\hat{a}$ is the annihilation operator, $[\hat{a},\hat{a}^{\dag}] = 1$. If we consider only one output port of the BS, we have the following evolution for the input signals: $\ket{\beta} \to |2\sqrt{\tau}\, \beta\rangle = \ket{\psi_1}$ and $\ket{-\beta} \to | 0 \rangle = \ket{\psi_0}$, respectively.
Under these conditions, the optimal strategy turns out to be the discrimination between the presence and the absence of light, which corresponds to the positive operator-valued measure $\{ \Pi_1=\sum_{n>0}\ketbra{n}{n}, \Pi_0=\ketbra{0}{0}\}$. It is worth noting that in this case the use of on/off photodetectors, without any photon-number discrimination power, is sufficient. In the limit $\tau \to 1$, the error probability $P_e = \frac12 ( \bra{\psi_1} \Pi_0 \ket{\psi_1} + \bra{\psi_0} \Pi_1 \ket{\psi_0} )$ in the discrimination is given by $P_e = \frac12 \exp(-4 |\beta|^2)$, which is twice the minimum error probability given by the Helstrom bound when $|\beta|\gg 1$ \cite{hel:76}.
\par
The previous treatment is based on the assumption that the relative phase $\phi$ between the input signal and the LO is constant and precisely known (in the present case $\phi=0$).
On the other hand, by setting $\alpha = | \beta | \, \e^{\imm\phi} \tau/(1- \tau)$ and still in the limit $\tau\to1$, the error probability reads $P_e(\phi) = \big [1 - {\rm e}^{ -4|\beta|^2\sin^2(\phi/2)} + {\rm e}^{ -4 |\beta|^2 \cos^2(\phi/2)}\big ] /2 $. Note that $P_e(\phi)\geq P_e(0) \equiv P_e$.
Furthermore,  for small values of $\phi$ and high signal energy $|\beta|^2$ we have $P_e(\phi) \approx P_e(0) + \left[\frac12 + P_e(0)\right] \, |\beta|^2\, \phi^2$: in this regime, the larger the energy, the larger the error probability.
\par
In our scheme, we assume that the input signal $\hrho(\beta)$ in Eq.~(\ref{input}) and the LO $\ket{\alpha\, \e^{\imm\phi}}$ ($\alpha, \beta \in \mathbbm R$ and $\alpha,\beta >0$) are mixed at a BS with transmittance $\tau$ \cite{note}. By defining $a = \alpha \sqrt{1-\tau}$  and $b = \beta \sqrt{\tau}$, the overall output state can be cast in the form $\hrho_{\text{out}}(a,b,\phi)=\hat{D}(a\, \e^{\imm\phi})\hrho(b)\hat{D}^\dag(a\, \e^{\imm\phi})$ and the photon-number distribution $\Tr{\hrho_{\text{out}}(a,b,\phi)\ketbra{n}{n}}$ of the transmitted beam reads
\begin{equation}\label{Pn}
p_n(a,b,\phi) =\frac{1}{2} \left( \e^{-\nu_+}\frac{\nu_+^n}{n!}+ \e^{-\nu_-}\frac{\nu_-^n}{n!} \right),
\end{equation}
which is the sum of two Poisson distributions depending on $\phi$ through the mean values $\nu_\pm=a^2+b^2\pm2 \,a\, b \cos\phi$. Therefore, the real-time monitoring of the phase is attainable directly from Eq.~(\ref{Pn}) by using the same acquired data and a suitable estimation strategy, as described in the following.

\par
\textit{Bayesian estimation} -- Let us assume that  $M$ signals are received by means of a PNR detector and $\{n_k\}=\{n_1, n_2, \ldots, n_M\}$, $n_k \in \mathbb{N}$, $\forall k$, is the data sample corresponding to the detected number of photons. As $\{n_k\}$ implicitly depends on $\phi$, we can build the sample probability $P(\{ n_k \} | \phi)=\prod_{n=0}^\infty  p_{n}(a,b,\phi)^{m_n}$, being $m_n$ the number of occurrences of $n$ detected photons, so that $\sum_n m_n=M$. $P(\{ n_k \} | \phi)$ is thus the probability of obtaining the actual sample given $\phi$. Thanks to Bayes theorem, we can write the posterior probability of $\phi$ given the sample, namely, $\Ppnr(\phi | \{ n_k \}) = \Npnr\, P(\{ n_k \} | \phi)$, where $\Npnr$ is a normalization factor and $\phi$ is assumed to be described by a uniform prior distribution. The Bayes estimator of $\phi$ is $\bar{\phi}=\int d\phi \,\phi\, \Ppnr(\phi | \{ n_k \}) $ and its variance is $\text{Var}_\phi=\int d\phi \,(\phi-\bar{\phi})^2\, \Ppnr(\phi | \{ n_k \})$.
It is well known \cite{bayes:1,bayes:2} that the Bayes estimator is asymptotically optimal if $M\gg 1$, that is $\text{Var}_{\phi} \to [M\, F^{\rm{\footnotesize{PNR}}}_{\phi}]^{-1}$, where $F^{\rm{\footnotesize{PNR}}}_{\phi}=\sum_n p_n(a,b,\phi) [\partial_\phi \ln p_n(a,b,\phi)]^2$ is the FI associated with $p_n(a,b,\phi)$.
\par
\begin{figure}[t!]
\includegraphics[width=0.23\textwidth]{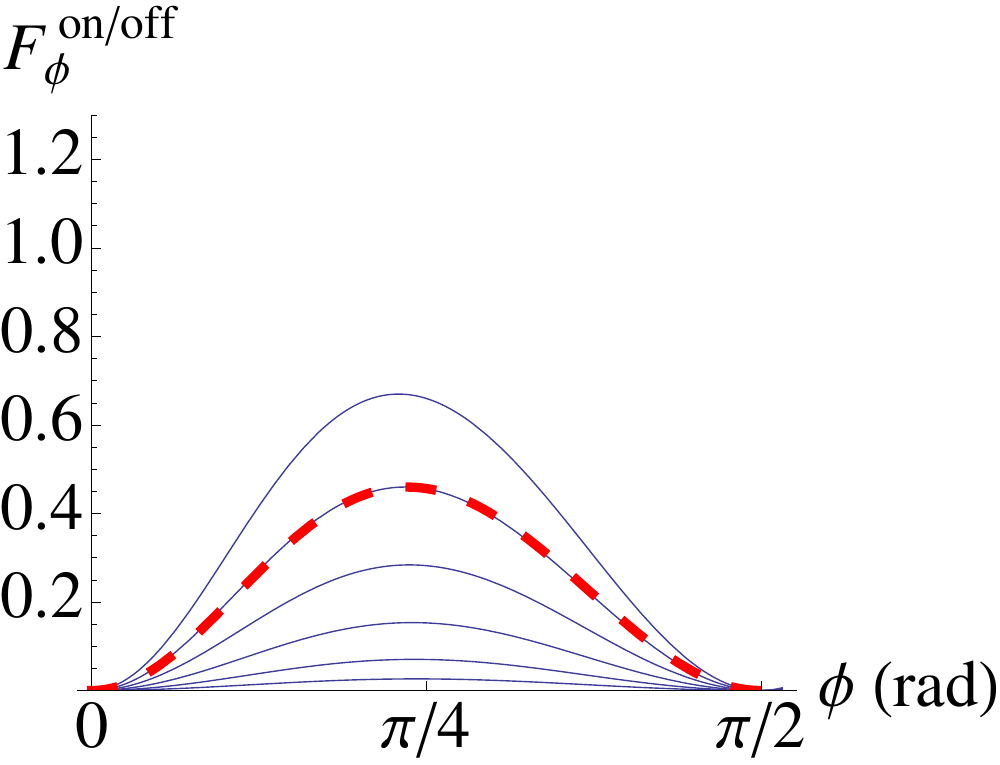}
\includegraphics[width=0.23\textwidth]{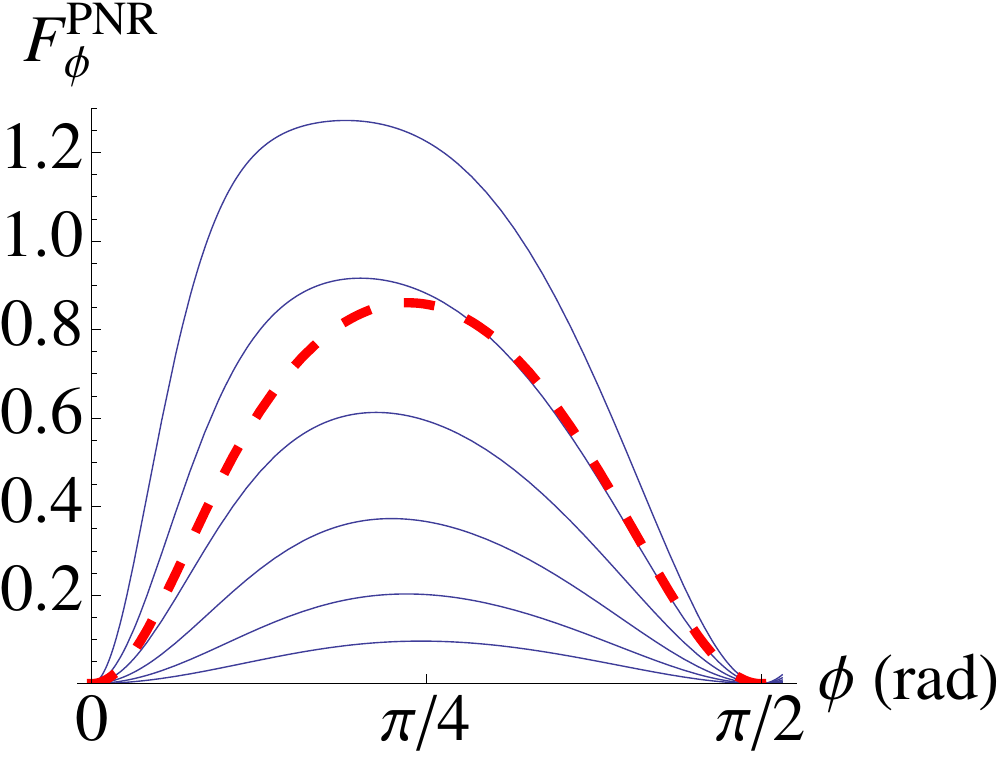}
\caption{(Color online) FI for on/off (left) and PNR (right) detection as a function of $\phi$ for $a=b=0.5,\ 0.6.\ 0.7,\ 0.8,\ 0.9, 1.0$  (blue lines, from bottom to top). In the two panels, the same encoding has been used in order to emphasize the higher values of $F^{\rm{\footnotesize{PNR}}}_\phi$ with respect to $F^{\text{on/off}}_\phi$. The red dashed lines are the FIs as expected in our experimental configuration ($a=1.12$ and $b=0.79$).
}
\label{f:FisherMaxima}
\end{figure}
Since on/off detection can be seen as a particular case of PNR detection, the Bayes estimator can now be obtained straightforwardly. We identify the probabilities $P_{\text{off}}\equiv p_0(a,b,\phi)$ and $P_{\text{on}}\equiv \sum_{n>0} p_n(a,b,\phi)=1-P_{\text{off}}$. Therefore, given the same sample $\{n_k\}$ considered above, the sample probability reduces to $P(\{ n_k \} | \phi)=P_{\text{off}}^{m_{\text{off}}} P_{\text{on}}^{m_{\text{on}}}$, where $m_{\text{off}}$ and $m_{\text{on}}=M-m_{\text{off}}$ are the number of  ``off'' ($n_k=0$) and ``on''  ($n_k>0$) events, respectively. The posterior probability is thus given by $P_{\text{on/off}}(\phi | \{ n_k \}) = N_{\text{on/off}}\, P(\{ n_k \} | \phi)$. Also in this case the Bayes estimator asymptotically reaches the optimal value, where the corresponding FI reduces to $F^{\text{on/off}}_\phi=(\partial_\phi P_{\text{off}})^2 (P_{\text{off}} P_{\text{on}})^{-1}$. In Fig.~\ref{f:FisherMaxima} we report the behavior of the FI in the case of on/off e PNR detection for a particular choice of the involved parameters. The FI displays a maximum in the interval $(0,\pi/2)$, whereas it vanishes at $\phi=0,\pi/2$ \cite{note:pi:4}.
\par
\begin{figure}
\includegraphics[width=0.37\textwidth]{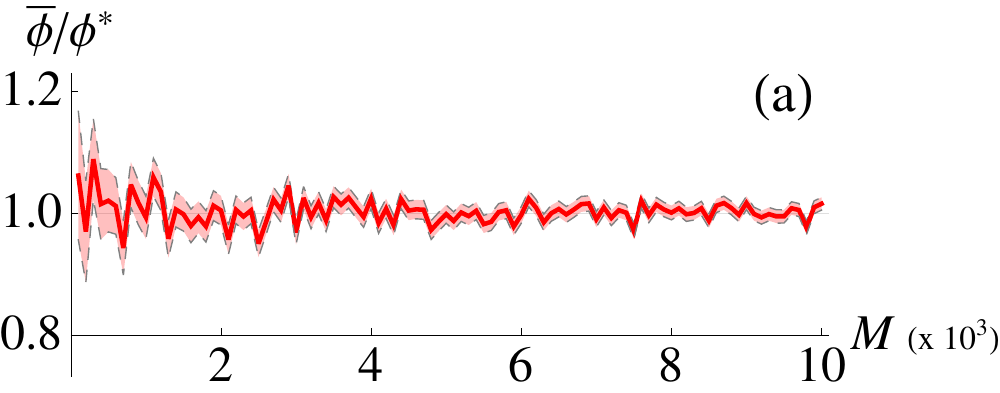}\\
\includegraphics[width=0.37\textwidth]{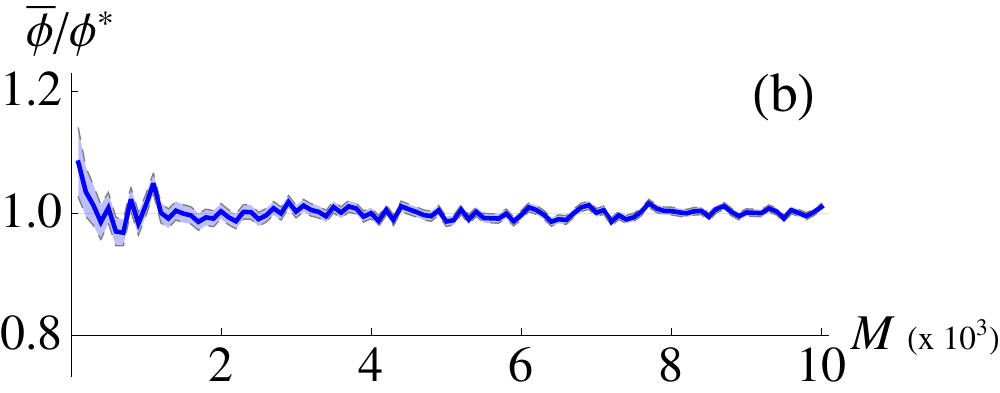}\\
\includegraphics[width=0.37\textwidth]{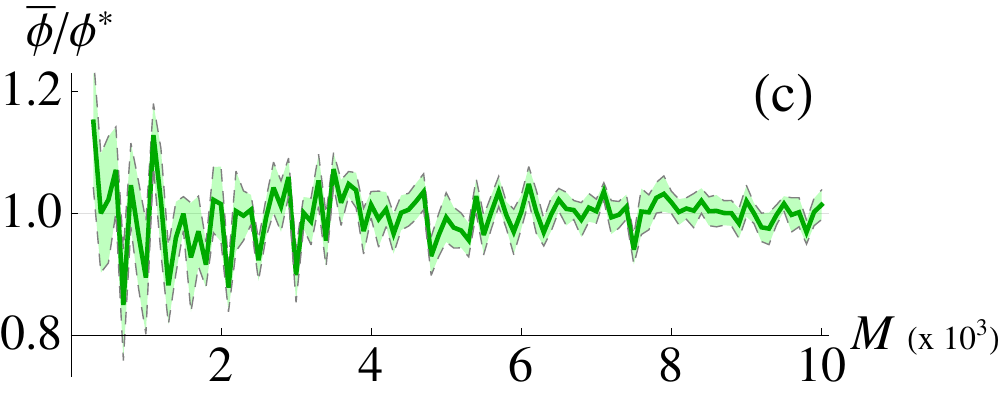}
\caption{(Color online) Phase estimation with simulated data using different methods: Bayesian strategies with on/off detectors (a) and PNR detectors (b), and the inversion of the Fano factor (c). The plots show the ratio $\bar{\phi}/\phi^*$ (solid curve) and the corresponding standard deviation (dashed curves) as a function of the number $M$ of data sample. We set $a=b=\sqrt{2}$ and $\phi^*=0.3$. }
\label{f:BayesEstimate}
\end{figure}

\textit{Numerical simulations} -- In order to test the proposed approach for retrieving the relative phase, we firstly performed Monte Carlo simulated experiments. The data samples were generated according to the photon-number probability distribution in Eq.~(\ref{Pn}), corresponding to the expected output state $\hrho_{\text{out}}(a,b,\phi^*)$, where $\phi^*$ is the actual value of the phase. In our simulations we chose to set $\phi^* = 0.3 $. We applied the Bayesian method to both the cases of on/off and PNR detection of the output signal and compared the two corresponding estimators $\bar{\phi}$ and their standard deviations [see Fig.~\ref{f:BayesEstimate}~(a) and (b)]. As it is apparent from the figure, the employment of PNR detectors brings the estimator $\bar{\phi}$ to converge more rapidly, just after $M\sim 10^3$, to the expected value $\phi^*=0.3 $ than using the on/off scheme.
\par
The Bayesian method is very powerful as it converges fast and, as we will see later, it provides a result robust against phase noise. Nonetheless, information about the phase parameter can also be retrieved in other ways, such as by inverting some measurable quantities depending on the parameter. Regarding our scheme, the measurement of the Fano factor $\mathcal{F}(\phi)=\text{Var}[\hat{N}]/\langle \hat{N} \rangle$ of the output state at the BS can be easily implemented by reconstructing the photon statistics of the output by means of PNR detectors. The Fano factor of the displaced state $\hrho_{\text{out}}(a,b,\phi)$ considered beforehand displays an explicit dependence on the phase parameter \cite{AOB:IQIS13} that can be inverted to obtain $\phi$. By analyzing the same Monte Carlo simulated data used before, according to the photon statistics $p_n(a,b,\phi)$, we can compare this inversion method with the Bayesian one. In particular, in Fig.~\ref{f:BayesEstimate}~(c) we show the estimation of $\phi$ and the corresponding standard deviation as functions of the number of data $M$, with the same parameters employed for the Bayesian estimation. The plot clearly shows a slower convergence to the expected value $\phi^*$, with very large fluctuations due to error propagation in the inversion method.
\par
\begin{figure}
\includegraphics[width=0.35\textwidth]{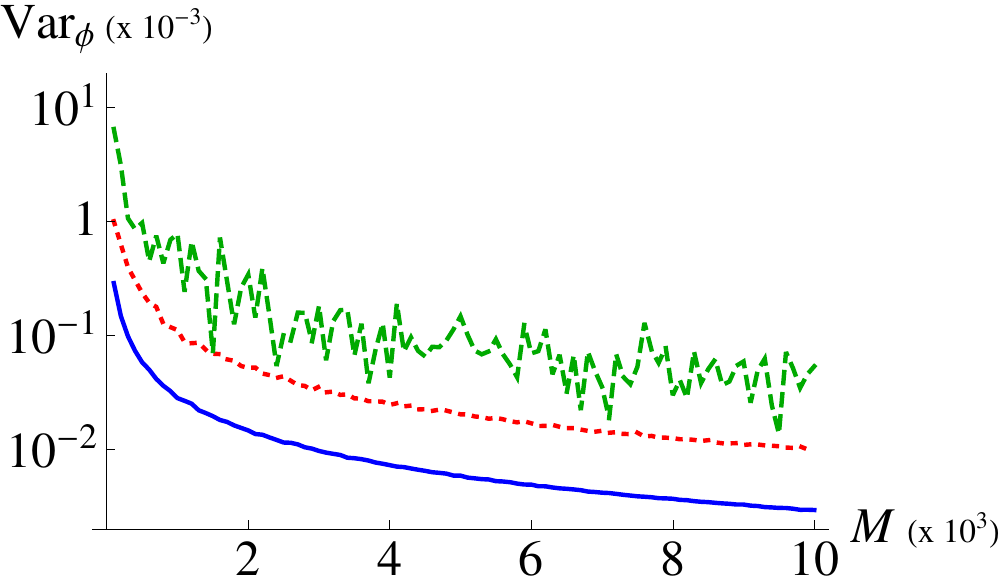}
\caption{\label{f:Var}(Color online) Logarithmic plot of the variance $\text{Var}_\phi$, in numerical simulations, given by the Bayesian method for on/off detectors (red, dotted) and PNR detectors (blue, solid), and by the method of inversion of the Fano factor (green, dashed). We set $\phi^*=0.3$ and $a=b=\sqrt{2}$.}
\end{figure}
In Fig.~\ref{f:Var} we compare the variances, plotted against the number of data samples $M$, provided by the three methods used for retrieving the phase parameter $\phi$: the Bayesian strategy for on/off and PNR detectors and the method of inversion of the Fano factor. The convergence to the expected value $\phi^*$ is clearly faster for Bayesian strategies than for the inversion of the Fano factor. In particular, the employment of PNR detectors results the best strategy since more information can be extracted from the reconstruction of the photon statistics.
\par
In a more realistic scenario the input coherent states can be affected by noise during propagation. In this Letter  we model the noise by introducing a uniform phase noise \cite{note:Ph:noise}. The resulting state is the so-called bracket state $\hrho(b,\gamma)\equiv \gamma^{-1}\int_{-\gamma/2}^{\gamma/2} d \psi\, \hrho(b\, \e^{\imm\psi})$ \cite{AOB:IQIS13}, where $\gamma$ is the noise parameter.
\begin{figure}
\includegraphics[width=0.35\textwidth]{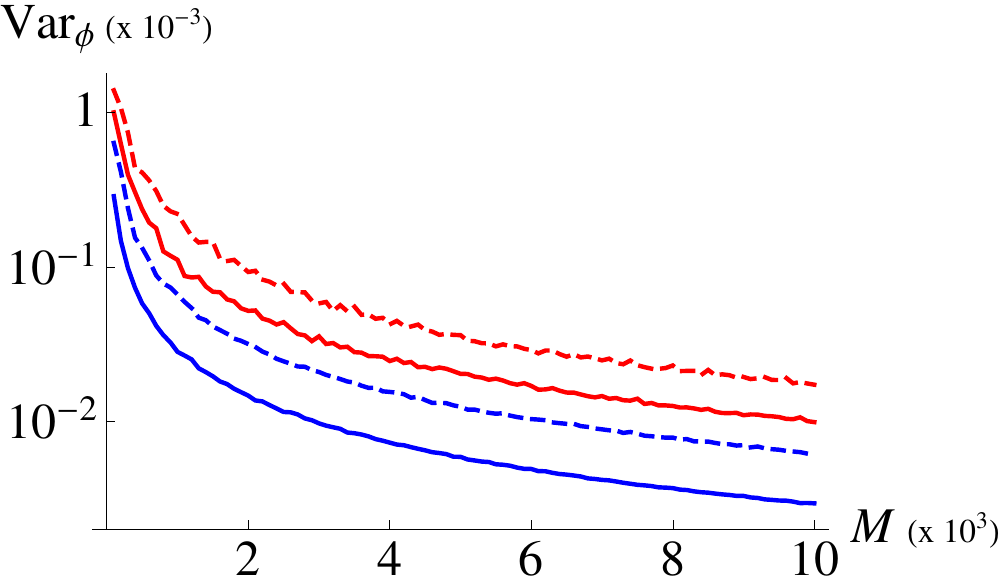}
\caption{\label{f:VarGamma}(Color online) Logarithmic plot of the variance $\text{Var}_\phi$, in numerical simulations with input bracket states $\hrho(b,\gamma)$, given by the Bayesian method for on/off detectors (red, dashed) and PNR detectors (blue, dashed), compared to the ideal case $\gamma\to 0$ (solid lines). Simulated setup parameters:  $\phi^*=0.3$, $a=b=\sqrt{2}$ and $\gamma=\pi/4$.}
\end{figure}
Our aim is now to apply the same Bayesian approach by considering the photon-number probability distribution $p_n(a,b,\phi,\gamma)\equiv \gamma^{-1 }\int_{-\gamma/2}^{\gamma/2}d \psi\, p_n(a,b,\phi-\psi)$ corresponding to the output state $\hrho_{\text{out}}(a,b,\phi,\gamma)=\hat{D}(a\, \e^{\imm\phi})\hrho(b,\gamma)\hat{D}^\dag(a\, \e^{\imm\phi})$. The Bayesian strategies prove to be very robust in the phase estimation with this kind of noise, only showing small differences in the convergence (see Fig. \ref{f:VarGamma}) compared to the ideal case in the limit of vanishing noise ($\gamma\to 0$). As expected, the effect of the uniform phase noise described by the bracket states is a slight increase of the variance of the estimation procedure.
\par
\begin{figure}[t]
\includegraphics[width=0.22\textwidth]{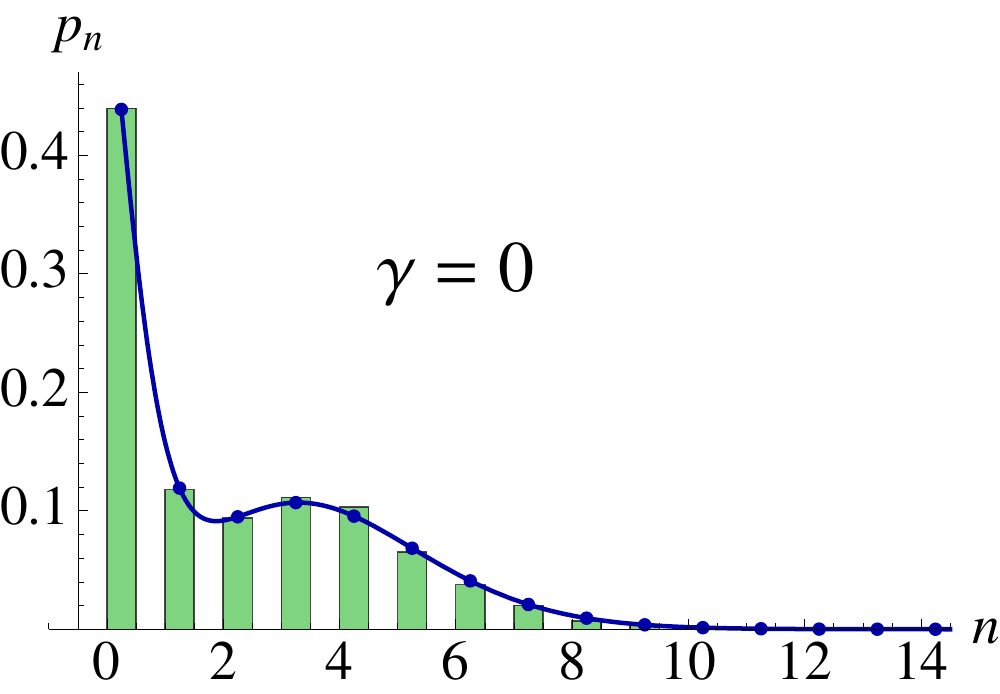}
\includegraphics[width=0.25\textwidth]{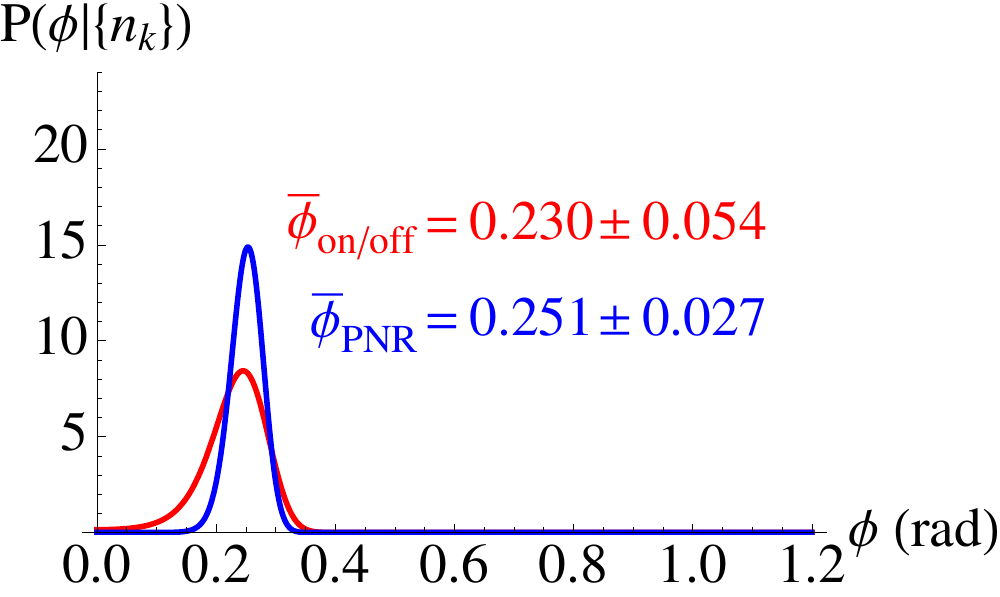}\\
\includegraphics[width=0.22\textwidth]{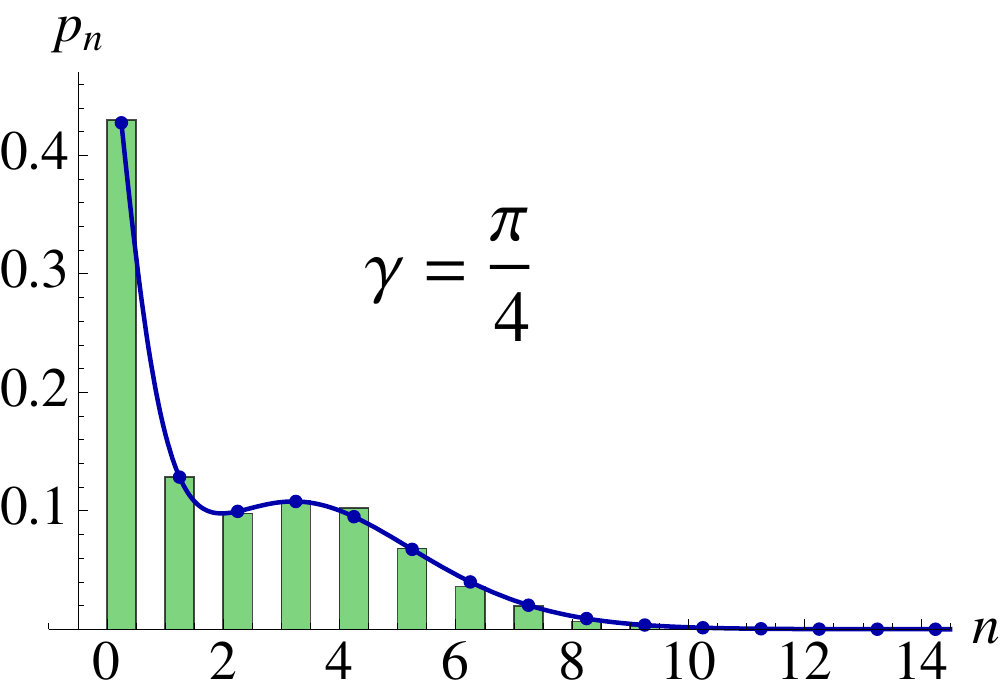}
\includegraphics[width=0.25\textwidth]{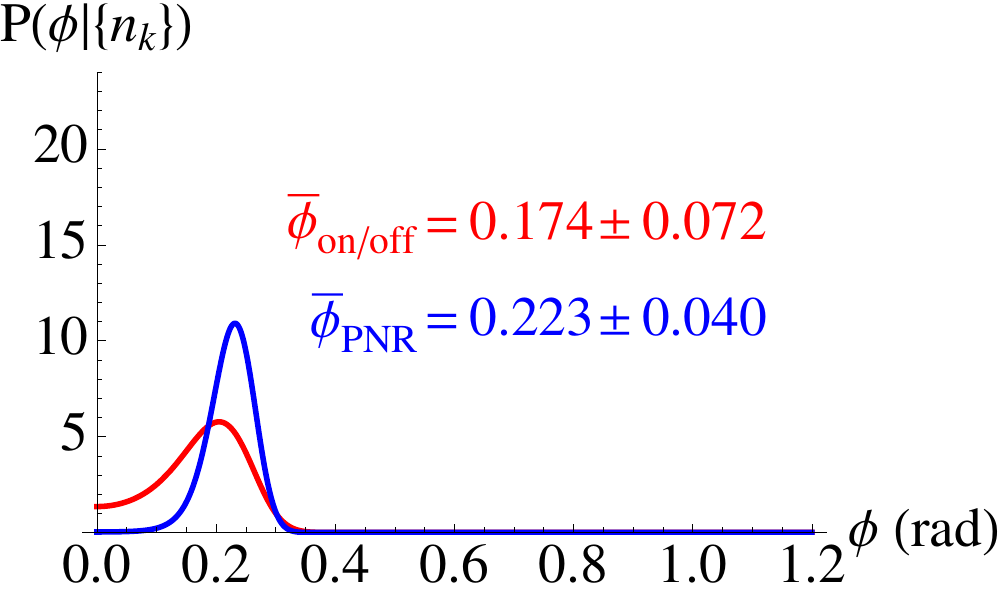}
\caption{(Color online) Left: experimental photon-number distributions $p_n^{\rm (exp)}$  obtained by PNR detection (green histograms) and theoretical expectations $p_n \big ( a,b,\bar{\phi}_{\rm PNR},\gamma \big )$ (blue lines) . Right: Bayesian probability distributions for the parameter $\phi$, corresponding to a set of $M=4\,000$ experimental data acquired with both on/off (red curve) and PNR (blue curve) detectors.
The experimental parameters are $a=1.12$, $b=0.79$, $\phi^* \simeq 0.25$, $\gamma=0$ (top row) and $\gamma=\pi/4$ (bottom row). }
\label{f:Exp:Bayes:Pn:1}
\end{figure}
\textit{Experimental results} --
In our experimental setup, linearly-polarized pulses at 523~nm (5-ps-pulse duration, 500-Hz-repetition rate) were sent to a Mach-Zehnder interferometer, in which the relative phase between the two arms was changed by means of a piezoelectric movement operated step by step. The output state was delivered to a hybrid photodetector (HPD, R10467U-40, Hamamatsu; maximum quantum efficiency $\sim$ 0.5 at 500~nm, 1.4-ns response time), amplified and synchronously integrated \cite{pnr:1,pnr:2}. By exploiting the linearity of our device, the value of the relative phase can be determined at each piezo position, independent of the regularity and reproducibility of the movement \cite{JosaBwigner}. Bracket states can be obtained in post-selection according to $\phi^*$ and the noise parameter $\gamma$.
More in detail, the bracket states have been produced by selecting data samples corresponding to two different choices of the relative phase, namely $\phi^* \simeq 0.25$ and $\phi^* \simeq \pi/4$ with $\gamma=\pi/4$ and $\gamma=\pi/2$, respectively.
\par
\begin{figure}[t]
\includegraphics[width=0.22\textwidth]{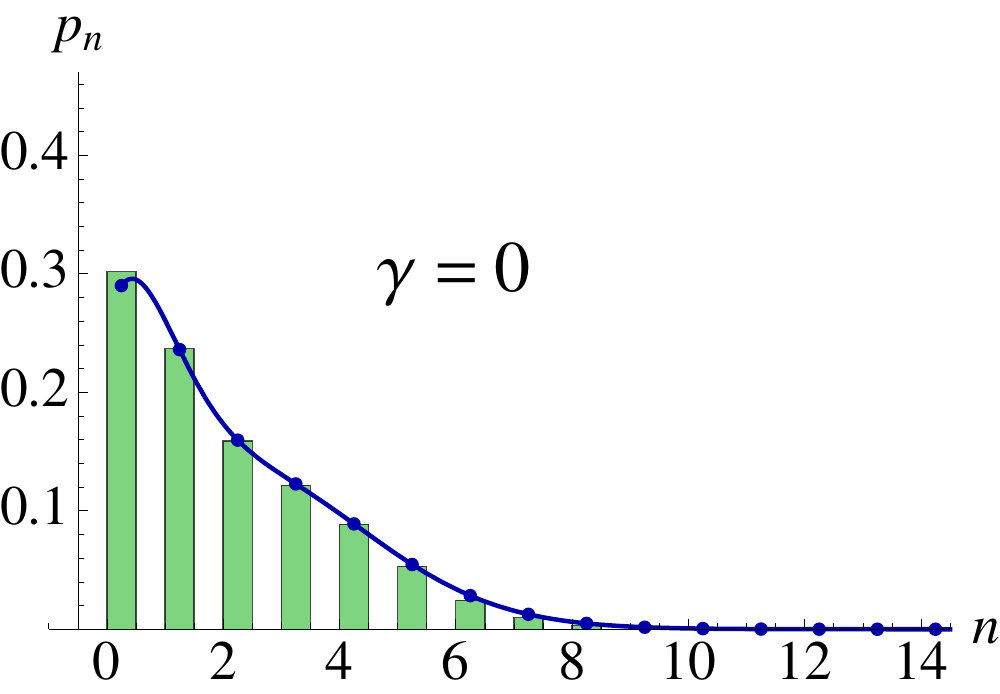}
\includegraphics[width=0.25\textwidth]{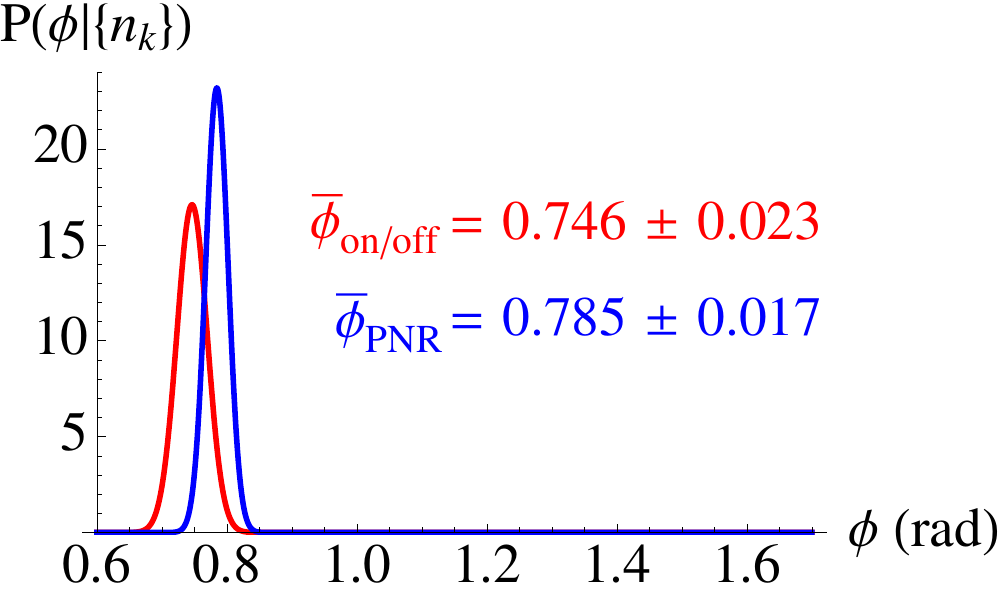}\\
\includegraphics[width=0.22\textwidth]{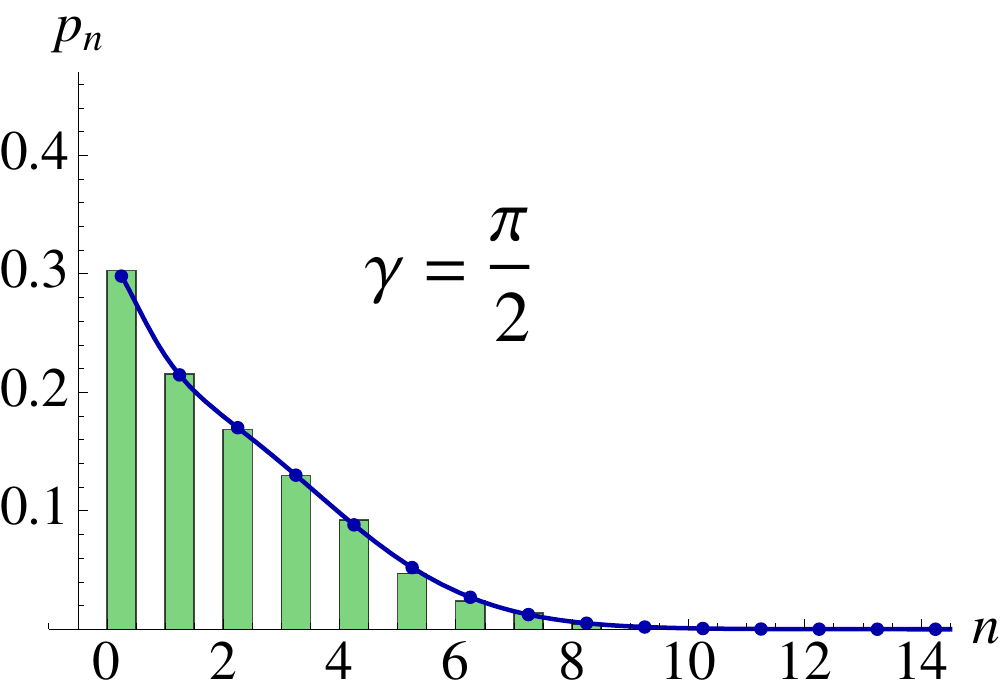}
\includegraphics[width=0.25\textwidth]{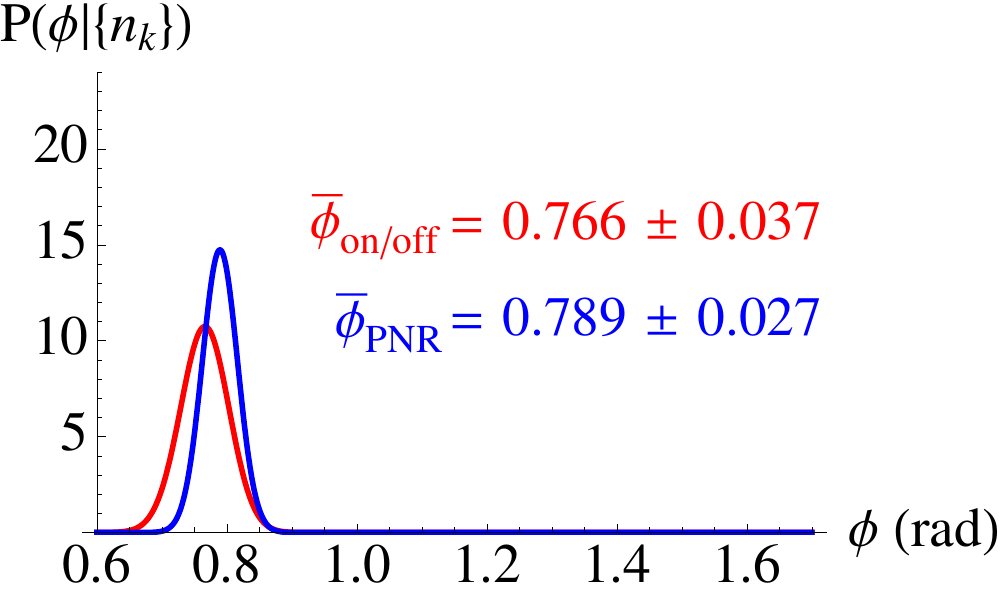}
\caption{(Color online) Left: experimental photon-number distributions $p_n^{\rm (exp)}$ obtained by PNR detection (green histograms) and theoretical expectations $p_n \big ( a,b,\bar{\phi}_{\rm PNR},\gamma \big )$ (blue lines). Right: Bayesian probability distributions for the parameter $\phi$, corresponding to a set of $M=4\,000$ experimental data acquired with both on/off (red curve) and PNR (blue curve) detectors. The experimental parameters are $a=1.12$, $b=0.79$, $\phi^* \simeq \pi/4$, $\gamma=0$ (top row) and $\gamma=\pi/2$ (bottom row).}
\label{f:Exp:Bayes:Pn:2}
\end{figure}

The results are presented in Figs.~\ref{f:Exp:Bayes:Pn:1} and \ref{f:Exp:Bayes:Pn:2}, where we plot the measured photon statistics $p_n^{\rm (exp)}$, the posterior Bayesian distributions $P_{\text{on/off}}(\phi | \{ n_k \})$ and $\Ppnr ( \phi | \{ n_k \})$ and the corresponding estimated phases  $\bar{\phi}_{\text{on/off}}$ and $\bar{\phi}_{\rm PNR}$ with their variances. Furthermore, in order to assess the quality of our reconstructions, we also plot the photon distributions $p_n \big ( a,b,\bar{\phi}_{\rm PNR},\gamma \big )$,  which have fidelities $F=\sum_n \sqrt{ p_n \big ( a,b,\bar{\phi},\gamma \big )\, p_n^{{\rm (exp)}} }$ higher than $99.9\%$ to the experimental data \cite{BinaFid}. In both Figures it is evident that the employment of a PNR detector allowed us to reconstruct with accuracy the photon number distribution $p_n \big ( a,b,\bar{\phi},\gamma \big )$ for the two bracket states (see left panels of Figs.~\ref{f:Exp:Bayes:Pn:1} and \ref{f:Exp:Bayes:Pn:2}) and to apply the Bayesian method to both the cases of on/off and PNR detection. If we focus on the first experimental test concerning  $\phi^*\simeq 0.25$ (Fig.~\ref{f:Exp:Bayes:Pn:1}), we can see that the Bayesian probability distributions, affected by a bias, show an asymmetric shape which is more marked in the case of on/off detection (see right panels in  Fig.~\ref{f:Exp:Bayes:Pn:1}) \cite{Moda}. Strikingly, PNR detection enhances the estimation as the Bayesian distribution is more peaked, has a smaller variance and a reduced asymmetry.
The second experimental test for $\phi^*\simeq \pi/4$ displays better performances (see right panels in  Fig.~\ref{f:Exp:Bayes:Pn:2}), in agreement with the behavior of FI in the plots of Fig.~\ref{f:FisherMaxima} (red dashed lines). PNR detection, again, provides an enhancement in the phase estimation describable in terms of a smaller variance and a more peaked Bayesian posterior distribution. In both situations, we note that the effect of the phase noise is to broaden the Bayesian distributions by increasing the variance, in agreement with the simulated data and the plot in Fig.~\ref{f:VarGamma}. Nonetheless, the Bayesian strategy turns out to be very robust in the presence of phase noise, even if it is modeled, as in the present analysis, by a uniform ``white-noise'' with a wide range such as $\gamma=\pi/4$ (second row of Fig.~\ref{f:Exp:Bayes:Pn:1}) and $\gamma=\pi/2$ (second row of Fig.~\ref{f:Exp:Bayes:Pn:2}).
\par
\textit{Conclusions} -- We have proposed and demonstrated
a real-time method to monitor the phase reference of a Kennedy-like receiver
without stopping the communication.  We have provided the experimental realization
of the protocol, which strengthen our theoretical model and numerical simulations.
Phase estimation is usually implemented with a pure, optimized
input probe state. Nevertheless, here we have shown that it is possible to perform
a useful phase estimation also with a mixed, not optimized, probe state,
that is the overall mixture of the two coherent states encoding the signal.
Our strategy is based on Bayesian analysis and thus allows us to asymptotically reach the
minimum uncertainty in the estimation collecting just  few
thousands of data.  Furthermore, we have demonstrated the advantages of using
PNR detectors with respect to on/off detectors and we have tested the performance of
our strategy also in the presence of uniform phase noise.
Our results pave the way to the real-time monitoring of the phase reference
in more complicated communication systems or quantum optics setups,
which require a precise control of the phase, e.g. in the generation of
nonclassical states, such as squeezed states.

\textit{Acknowledgments} -- This work has been supported by MIUR (FIRB ``LiCHIS'' Ñ RBFR10YQ3H).


\begin{thebibliography}{30}
\bibitem{note}Actually, if $\alpha = \beta$, the configuration with $\tau = 1/2$ corresponds to the quasi optimal discrimination between the coherent states $\ket{\pm \beta / \sqrt{2}}$.
\bibitem{note:pi:4} It is worth noting that in both on/off and PNR detection, the FI goes to zero as $\phi \to 0$: this is a consequence of the probe mixed state $\hrho(\beta)$ of Eq.~(\ref{input}), which is a balanced mixture of the state $\ket{\beta}$, that maximizes the FI for $\phi = 0$, and $\ket{-\beta}$, that leads to a null FI for $\phi = 0$.
\bibitem{note:Ph:noise} We do not consider losses since they only affect the amplitude of the
coherent states but leave unchanged the other physical properties, such as the photon statistics,
that is simply rescaled, and the phase.
\bibitem{Moda} In the case of small phase parameter, $\phi\simeq 0.25$, the optimal estimator turns out to be the mode $\tilde{\phi}\equiv \text{Max}\{ P(\phi | \{ n_k \}) \}$, as the Bayes posterior probabilities are biased, resulting in $\tilde{\phi}_{\text{on/off}}=0.24$ and $\tilde{\phi}_{\text{PNR}}=0.25$ (for $\gamma=0$) and $\tilde{\phi}_{\text{on/off}}=0.20$ and $\tilde{\phi}_{\text{PNR}}=0.23$ (for $\gamma=\pi/4$).
\end{thebibliography}
\end{document}